\documentclass[fleqn,usenatbib]{mnras}

\usepackage{newtxtext,newtxmath}

\usepackage[T1]{fontenc}

\DeclareRobustCommand{\VAN}[3]{#2}
\let\VANthebibliography\thebibliography
\def\thebibliography{\DeclareRobustCommand{\VAN}[3]{##3}\VANthebibliography}

\usepackage{graphicx}	
\usepackage{amsmath}	
\usepackage{amssymb}	

\title[Bright third star around contact binary]{The new compact triple system: discovery of bright third star around contact binary using LAMOST-MRS spectra and photometry. }

\author[M. Kovalev et al.]{
Mikhail Kovalev,$^{1,2,3}$\thanks{E-mail: mikhail.kovalev@ynao.ac.cn}
Azizbek Matekov,$^{1,4,5}$
Sufen Guo,$^{6,1,2,3}$
Xuefei Chen$^{1,2,7}$
\newauthor
and Zhanwen Han$^{1,2,7}$\\
$^{1}$Yunnan Observatories, China Academy of Sciences, Kunming 650216, China\\
$^{2}$Key Laboratory for the Structure and Evolution of Celestial Objects, Chinese Academy of Sciences, Kunming 650011, China\\
$^{3}$International Centre of Supernovae, Yunnan Key Laboratory, Kunming 650216, China\\
$^{4}$ University of Chinese Academy of Sciences, Yuquan Road 19, Beijing 100049 Sijingshang Block, China\\
$^{5}$Ulugh Beg Astronomical Institute, Uzbekistan Academy of Sciences, 33 Astronomicheskaya str., Tashkent, 100052, Uzbekistan\\
$^{6}$School of Physical Science and Technology, Xinjiang University, Urumqi, 830046, China\\
$^{7}$Center for Astronomical Mega-Science, Chinese Academy of Sciences, 20A Datun Road, Chaoyang District, Beijing 100012, China\\
}

\date{Accepted XXX. Received YYY; in original form ZZZ}

\pubyear{2025}

\def\kms{\,{\rm km}\,{\rm s}^{-1}}

\def\feh{\hbox{[Fe/H]}}

\newcommand{\teff}{T_{\rm eff}}
\newcommand{\rv}{{\rm RV}}
\def\Vmic{V_{\rm mic}}

\def\vsini{V \sin{i}}
\def\logg{\log{\rm (g)}}
\def\snr{\hbox{S/N}}
\newcommand{\ha}{\hbox{H$\alpha$}}

\begin{document}
\label{firstpage}
\pagerange{\pageref{firstpage}--\pageref{lastpage}}
\maketitle

\begin{abstract}
We present a study of the third  star orbiting around  known contact eclipsing binary J04+25 using spectra from the LAMOST medium-resolution survey (MRS) and publicly available photometry. This is a  rare case of a hierarchical triple, where the third star is significantly brighter than the inner contact subsystem. We successfully extracted radial velocities for all three components, using the binary spectral model in two steps. Third  star radial velocities have high precision and allow direct fitting of the orbit. The low precision of radial velocity measurements in the contact system is compensated by large number statistics.
We employed a template matching technique for light curves to find periodic variation due to the light time travel effect (LTTE) using several photometric datasets. Joint fit of third  star radial velocities and LTTE allowed us to get a consistent orbital solution with $P_3=941.40\pm0.03$ day and $e_3=0.059\pm0.007$. We made estimations of the masses $M_{\rm 12,~3}\sin^3{i_3}=1.05\pm0.02,~0.90\pm0.02~M_\odot$ in a wide system and discussed possible determination of an astrometric orbit in the future data release of Gaia. Additionally, we propose an empirical method for measuring a period and minimal mass of contact systems, based on variation of the projected rotational velocity ($\vsini$) from the spectra.


\end{abstract}




\section{Introduction}
Contact binaries are a special type of eclipsing binaries, which have a distinct shape of the light curves (LC) and usually short periods. Thus they can be easily identified in photometrical surveys such as  Transatlantic Exoplanet Survey (TrES) \citep{tres}, Wide Angle Search for Planets (SuperWASP) \citep{swasp}, All Sky Automated Survey for SuperNovae (ASAS-SN) \citep{asassnv}, Zwicky Transient Facility (ZTF) \citep{ztf_var}, Asteroid Terrestrial-impact Last Alert System (ATLAS) \citep{atlas1,atlas2}, Kepler \citep{kepler} and Transiting Exoplanet Survey Satellite (TESS) \citep{tess}. There is much evidence that the formation of such systems will require interaction with the third body \citep{pribulla,dangelo,rucinski}, which can usually reveal itself through third light ($L3$) \citep{six_contact} or LTTE \citep{ltte}. These systems are actually hierarchical triples, containing the inner system with a short period and another third star on the wide, long-period orbit \citep{compacttriplesGDR3,newtriple}. 
\par
In this article we present  the discovery and detailed study of such a triple system:  J042901.09+254144.2, furthermore J04+25 (aka T-Tau0-03027; see other designations and basic info in Table~\ref{tab:des}), which was first discovered as an eclipsing binary in TrES data \citep{tres}. It was observed during the binary-related time-domain sub-survey (TD-B) of LAMOST-MRS \citep{mrs} and attracted our attention during the construction of the catalogue of spectroscopic orbits \citep{sb2cat}. We analyze available spectra and public photometry to confirm the presence of the third  star and provide a consistent orbital solution for it. J04+25 is a rare case when an inner contact system is orbited by the brighter third star with a relatively short period ($\sim3$ years), which allows us to measure precise masses.

\begin{table}
    \centering
    \caption{Designations and basic info from the literature: a-\protect\cite{gaia3}, b-\protect\citet{varstarindex}, c-\protect\cite{tic}, d-\protect\cite{asassnv}, e-\protect\cite{tres}, f-\protect\cite{cat23}, g-\protect\cite{apogee17}.}
    \begin{tabular}{lcc}
    \hline
    Property & Value & Reference\\
    \hline
    Gaia DR3 &    & a \\
    $\texttt{source\_id}$ & 151001777293928576\\
    $\alpha^\circ$ ICRS & 67.2545\\
    $\delta^\circ$ ICRS & 25.6955\\
    $\varpi$, mas & 1.6581$\pm$0.1175\\
    $\mu_\alpha \cos{\delta}$, mas year$^{-1}$ & 1.363$\pm$0.143 & \\
    $\mu_\delta$, mas year$^{-1}$ & -7.789$\pm$0.105 & \\
    $G$, mag & 13.000$\pm$0.005\\
    $\rv,\,\kms$ &-44.25$\pm$4.43\\
    \hline
    Variable Star indeX &  163229  & b \\
    TESS input catalogue & TIC268143030 & c\\
    ASAS-SN & J042901.02+254144.6	 & d\\
    TrES & T-Tau0-03027 &  e\\
    LAMOST & J042901.09+254144.2 & f\\
    APOGEE &2M04290108+2541443 & g\\
    
    
    \hline
    \end{tabular}
    \label{tab:des}
\end{table}

\par
The paper is organised as follows: in Section~\ref{sec:obs} we describe the observations. Section \ref{sec:methods} describes the methods and presents our results. In Section~\ref{discus} we discuss the results. In Section~\ref{concl} we summarise the paper and draw conclusions.
 Throughout the paper we indicate components of the inner contact system by indices ``1,2", while the third  star has index ``3". Inner orbit's parameters are referred to as ``12", while outer orbit related parameters have index ``3".

\section{Observations}
\label{sec:obs}
\subsection{Spectra}

LAMOST is a 4-meter quasi-meridian reflective Schmidt telescope with 4000 fibers installed on its $5\degr$ field of view focal plane. These configurations allow it to observe spectra for at most 4000 celestial objects simultaneously \citep{2012RAA....12.1197C, 2012RAA....12..723Z}.
 All available spectra were downloaded from \url{www.lamost.org/dr11/} under the designation J042901.09+254144.2. We use the spectra taken at a resolving power of $R=\lambda/ \Delta \lambda \sim 7\,500$. Each spectrum is divided into two arms: blue from 4950\,\AA~to 5350\,\AA~and red from 6300\,\AA~to 6800\,\AA. We convert the heliocentric wavelength scale in the observed spectra from vacuum to air using {\sc PyAstronomy} \citep{pyasl} and apply zero point correction using values from \cite{zb_rv}. Observations are carried out from 2018-01-25 till 2023-02-04, covering 18 nights with a time base of 1836 days.
 The period is short ($P_{12}=0.3642683$ d \cite{tres}), thus we analysed spectra taken during short 20 minute exposures, ignoring co-added spectra. In total, we have 73 spectra, where the average signal-to-noise ratio ($\snr$) of a spectrum ranges from 8 to 32 ${\rm pix}^{-1}$ for the blue arm and from 16 to 70 ${\rm pix}^{-1}$ for the red arm of the spectrum, with the majority of the spectra having $\snr$ around 40 ${\rm pix}^{-1}$.

\subsection{Photometry}
We checked publicly available photometric archives for J04+25 and found datasets in TrES, SuperWASP, ASAS-SN, ZTF, WISE, ATLAS, Gaia, Kepler, and TESS. It was also observed during the  Kilodegree Extremely Little Telescope (KELT) survey \citep{kelt}, but LC data are not available.

\begin{table*}
    \centering
    \begin{tabular}{cccccc}
         \hline
         LC & N & filter  & timebase, days & source & reference\\ 
         \hline
         SuperWASP & 6679 & $V$ & 1300 & \url{https://www.superwasp.org/} &
 \protect\citet{superwasp}\\
         TrES & 1171 & $V$ & 68 & VizieR & \cite{tres} \\
         ATLAS & 2376 & $o$ & 3372 & \url{https://fallingstar-data.com/forcedphot/queue/} & \cite{atlas3site}\\
               & 842  & $c$ & 3332 \\
        ZTF & 465 & $g$ & 2161 & \url{https://irsa.ipac.caltech.edu/} & \cite{ztf_main}\\
            & 681 & $r$ & 2041 \\
        ASAS-SN & 665 & $V$ & 2091 & \url{https://asas-sn.osu.edu/variables/206165} & \cite{asassnv}\\
            & 2091 & $g$ & 2504 \\
        WISE & 208 & $W1,W2$ & 4937 & \url{https://irsa.ipac.caltech.edu/} & \cite{wise}\\
        Gaia & 32 & $G$ & 919 & \url{https://www.cosmos.esa.int/gaia} & \cite{gaia3}\\
        Kepler K2 & 3430 & $K_p$ & 80 & MAST & \cite{k2mission}\\
        TESS 44 & 2845 & $T$ & 24 & MAST & \cite{tess}\\
        TESS 45 & 2994 & $T$ & 24 & \\
        TESS 70 & 9053 & $T$ & 24 & \\
        TESS 71 & 9256 & $T$ & 24 & \\
        \hline
    \end{tabular}
    \caption{Public photometry. VizieR for TrES \url{https://vizier.cds.unistra.fr/viz-bin/vizExec/Vgraph?J/AJ/135/850/T-Tau0-03027&P=0.36426830}, MAST \url{https://mast.stsci.edu/portal/Mashup/Clients/Mast/Portal.html} }
    \label{tab:lcs}
\end{table*}

Details on photometric datasets are provided in Table~\ref{tab:lcs}.
Only Quick Look Pipeline (QLP \citep{tess1,tess2}) LCs were available for 70,71 sectors, so we used QLP results for 44,45 sectors for consistency. 
 
\section{Methods \& Results} 
\label{sec:methods}

\subsection{Spectral fitting}

\label{sec:ind}
We analysed all spectra using the same single-star and binary model as in \cite{cat23}. Details on the synthetic spectral grid are provided in Appendix~\ref{sec:payne}. Here we provide a brief description and highlight updates which were necessary to fit spectra of the contact binary. 
The normalised binary model spectrum is generated as a sum of the two Doppler-shifted, normalised single-star model spectra ${f}_{\lambda,i}$ scaled according to the difference in luminosity, which is a function of the $\teff$ and visible stellar size. We use the following equation:    

\begin{align}
    {f}_{\lambda,{\rm binary}}=\frac{{f}_{\lambda,2} + k_\lambda {f}_{\lambda,1}}{1+k_\lambda},~
    k_\lambda= \frac{B_\lambda(\teff{_{,1}})}{B_\lambda(\teff{_{,2}})} k_R,
	\label{eq:bolzmann}
\end{align}
 where  $k_\lambda$ is the luminosity ratio per wavelength unit, $B_\lambda$ is the black-body radiation  (Planck function), $\teff$ is the effective temperature, $k_R$ - light ratio coefficient. Throughout the paper, we always assume the first component to be the brighter one.
\par
The binary model spectrum is later multiplied by the normalisation function, which is a linear combination of the first four Chebyshev polynomials \citep[similar to][]{Kovalev19}, defined separately for blue and red arms of the spectrum. The resulting spectrum is compared with the observed one using the \texttt{scipy.optimise.curve\_fit} function, which provides optimal spectral parameters, radial velocities (RV) of each component plus the light ratio and two sets of four coefficients of Chebyshev polynomials. We keep metallicity equal for both components. In total, we have 18 free parameters for a binary fit.

\begin{figure*}
    \includegraphics[width=\textwidth]{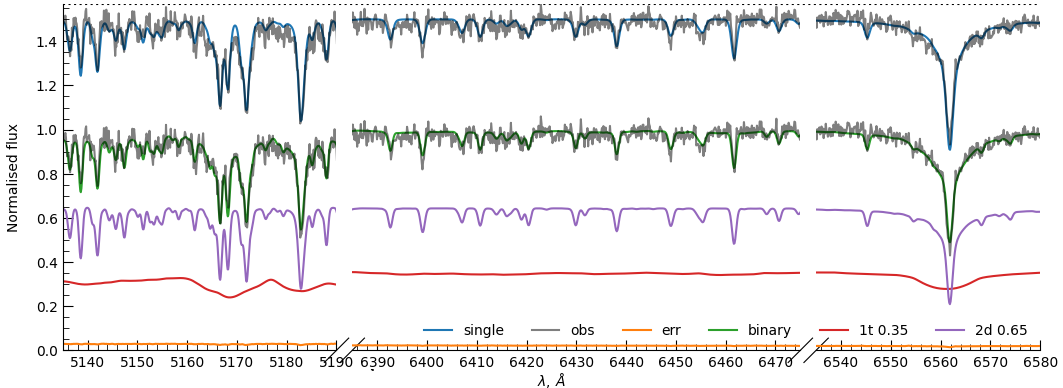}
	\includegraphics[width=\textwidth]{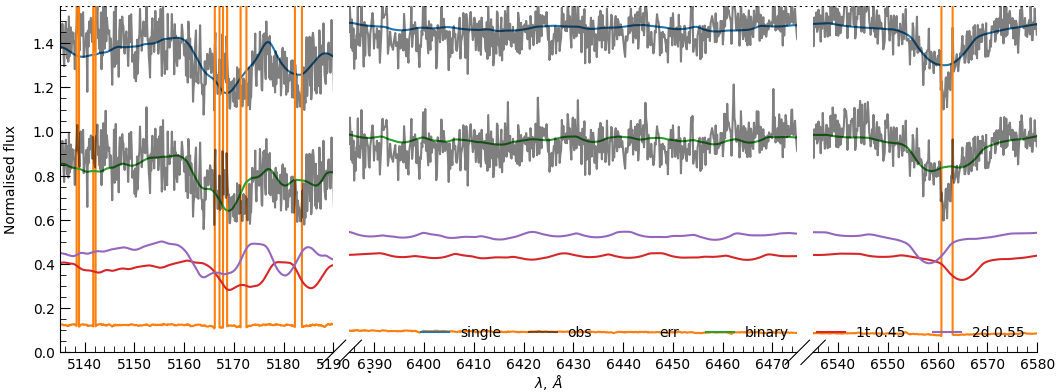}
    \caption{Example of the spectral fitting for spectrum taken at $\phi=0.25$, MJD=59190.669 d. Original spectrum (top) and residual spectrum after subtraction of narrow-lined component (bottom). We zoom into the wavelength range around the magnesium triplet, $\ha$ and in a 70~\AA~interval in the red arm. The observed spectrum is shown as a gray lines ( with different offsets), the best fits are shown as a green (binary model with zero offset) and blue (single-star model with +0.5 offset) lines. The primary component is shown as the magenta line, the secondary as a red line. Their light contribution is also shown in legend. The errors are shown as an orange line. }
    \label{fig:spfit}
\end{figure*}
\begin{figure}
    \includegraphics[width=\columnwidth]{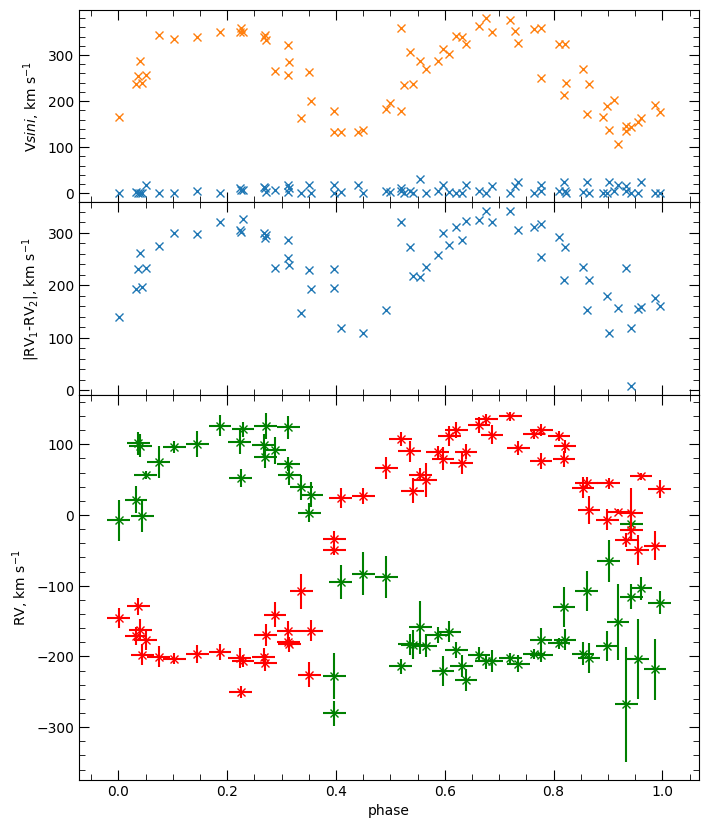}
    \caption{$\vsini$ of the dim spectral component (orange crosses) and bright spectral component (blue crosses) are shown in top panel as a function of orbital phase, computed using ephemeris from ASAS-SN. Middle panel shows absolute difference of $\rv$ fitted from residual spectrum. Bottom panel shows $\rv$, assigned to the both component of the contact system. Horizontal errorbars indicate half time of the exposure. }
    \label{fig:rv12}
\end{figure}

In the top panel of Figure~\ref{fig:spfit} we show the best fit examples by single-star and binary spectral models. We zoom into the wavelength range around the magnesium triplet, $\ha$ and in a 70~\AA~interval in the red arm, where many iron lines are clearly visible. Although the single-star model fit generally looks good as it captures all narrow spectral lines, it has a problem around Mg triplet lines (5160-5190Å). The binary model fits it well because it has an additional fast rotating component, which takes into account the spectral contribution of both components in the contact system. Derived radial velocities ($\rv$) for the primary allow us to get a single-line spectroscopic binary (SB1) orbit, see Section~\ref{sec:gls}. 
We can see that the primary spectral component contributes around 65\% in the spectrum, while the secondary component contributes to the remaining 35\%, although due to the variability of the inner contact binary, these contributions change from one spectrum to another. The first spectral component has narrow spectral lines and shows almost no change in spectral parameters, while parameters for the second spectral component change very quickly, since it corresponds to the contact binary. 
This system has a ``dumb-bell" shape, so the fitted value of projected rotational velocity ($\vsini$) will change according to the orbital phase computed using linear ephemeris from ASAS-SN \citep{asassnv}: 
\begin{align}
t_{min}(\rm HJD)=2457391.91388+0.364277E,
\label{eq:efimer}
\end{align}
where $E$ is an epoch (cycle number), see top panel in Figure~\ref{fig:rv12}. The bright spectral component has narrow spectral lines ($\vsini<20~\kms$), while the dim spectral component has $\vsini\sim150~\kms$ during eclipses and $\vsini\sim350~\kms$ at phases $\phi=0.25, 0.75$, when we can see the contact system from the side.  

In order to obtain the $\rv_{1,2}$ of the individual components in the contact system, we need to separate the bright spectrum of the third  star. This process is similar to opening the famous Russian doll ``Matryoshka". At first, we use the original fit to get the bright, narrow-lined component of the binary model. Once we remove it, we can fit the remaining spectrum with the binary model again. We should increase the error spectrum for the residual spectrum according to its relative contribution in the original spectrum. For strongest spectral lines, subtraction was not perfect; therefore, we also increased errors for them to infinity. Once we fitted all residual spectra, we removed results for ten of them with $\snr<30$ in the red arm, due to the bad performance of the binary model. We plot |$\rv_1-\rv_2$| as a function of orbital phase in the middle panel of the Figure~\ref{fig:rv12}. It is clear that |$\rv_1-\rv_2$| derived by the binary model correlates with $\vsini$ values derived by the binary model for the contact system in the previous step. However, $\rv$s need to be sorted because of confusion in the identification of primary/secondary by the binary model. Thus $\rv$ were folded in a phase diagram and assigned to each component based on orbital phase, see the bottom panel in the Figure~\ref{fig:rv12}. As expected for the contact binary, they follow a circular orbit pattern, with roughly equal amplitudes. The motion of the center of mass causes additional scatter in $\rv$, plus Rossiter-McLaughlin effect \citep{rossiter,mclaflin} changes $\rv$ near eclipse phases. It is clear that individual values for radial velocities have low precision, especially for spectra taken near conjunction phases; however, large number statistics (63 measurements) will allow us to use these data in the further analysis, together with photometric data. 
\par
We computed weighted averages of $\teff,\logg,\feh,\vsini$ for all three components using $\snr$ in the red arm as a weight. These results are collected in Table~\ref{tab:sp_par}. The third star is hotter than both components of the contact system and rotates much slower. Surface gravity $\logg_3=3.73\pm0.13$ dex, however, it can be underestimated when derived only from LAMOST-MRS spectra, as it was shown in \cite{j0647}. The third  star has subsolar metallicity $\feh_3=-0.24\pm0.05$ dex. Metallicity of the contact system is larger then for the third  star, although this is not very reliable result, since residual spectra are very noisy and no metal lines are visible due to significant rotational broadening. 

\begin{table}
    \centering
    \caption{Spectral parameters  for all three stellar components.}
    \begin{tabular}{lccc}
\hline
Parameter & Star 1 & Star 2 & Star 3 \\
\hline
$\teff$, K & 5637$\pm$342 & 5493$\pm$170 & 5847$\pm$38 \\
$\logg$, cgs & 4.13$\pm$0.41 & 3.96$\pm$0.46 & 3.73$\pm$0.13\\
$\feh$, dex & \multicolumn{2}{c}{$-0.06\pm0.13$} & -0.24$\pm$0.05\\
$\vsini,\,\kms$ & 162$\pm$36  & 155$\pm$39 & 6$\pm$8\\

\hline
    \end{tabular}
    \label{tab:sp_par}
\end{table}

\subsubsection{{\sc GLS} orbit of the third  star}
\label{sec:gls}
\begin{figure}
    \includegraphics[width=\columnwidth]{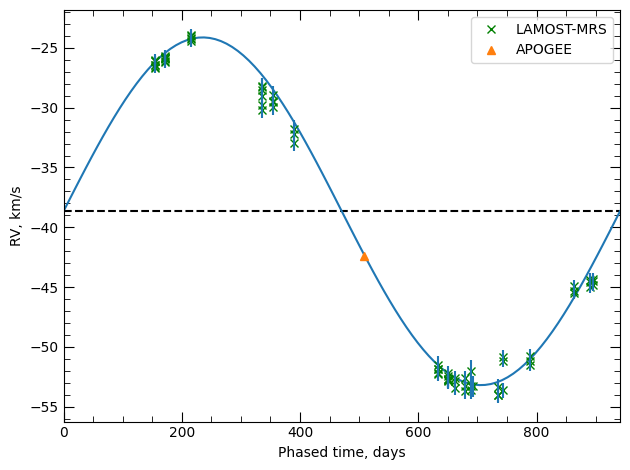}
    \caption{ {\sc GLS} orbital solution for the third star. Dashed horizontal line shows the systemic velocity. $\rv$ measurement from APOGEE DR17 \protect\cite{apogee17} is shown as an orange triangle.}
    \label{fig:gls}
\end{figure}

We can use radial velocities of the narrow-lined component,  which corresponds to the third  star, to calculate the orbital solution with Generalised Lomb-Scargle ({\sc GLS}) code \citep{gls}. We choose 73 datapoints from the first output of the binary spectral model and feed them to {\sc GLS}. Since the original {\sc GLS} doesn't provide uncertainties for eccentricity ($e_3$) and periastron angle ($\omega_3$) we refined the solution with the help of \texttt{scipy.optimise.curve\_fit} function. The resulting orbit is given in Figure~\ref{fig:gls} and Table~\ref{tab:rv_gls}.

\begin{table}
    \caption{ {\sc GLS} orbital solution for the third  star.}
    \centering
    \begin{tabular}{l|c}
    \hline
        parameter & value\\
    \hline
        $P_3$, day & 944.4146$\pm$5.7816 \\
        $t_{p3}$ , BMJD day & 58588.0089$\pm$29.0809\\
        $e_3$ & 0.038$\pm$0.017\\
        $\omega_3,\,^\circ$ & 23$\pm$12\\
        $K_3,\,\kms$ & 14.54$\pm$0.10\\
        $\gamma_3,\,\kms$ & -38.98$\pm$0.20\\
    \hline
    \end{tabular}
   
    \label{tab:rv_gls}
\end{table}

The third  star has a slightly eccentric orbit ($e_3=0.04\pm0.02$) with a relatively long period $P_3=944\pm5$ days, which is significantly longer than the period of the inner contact system. Our orbit agrees with a single $\rv$ measurement from APOGEE DR17 \citep{apogee17}.

\subsection{ Modelling with {\sc W-D}}

\begin{figure}
    \centering
    \includegraphics[width=\columnwidth]{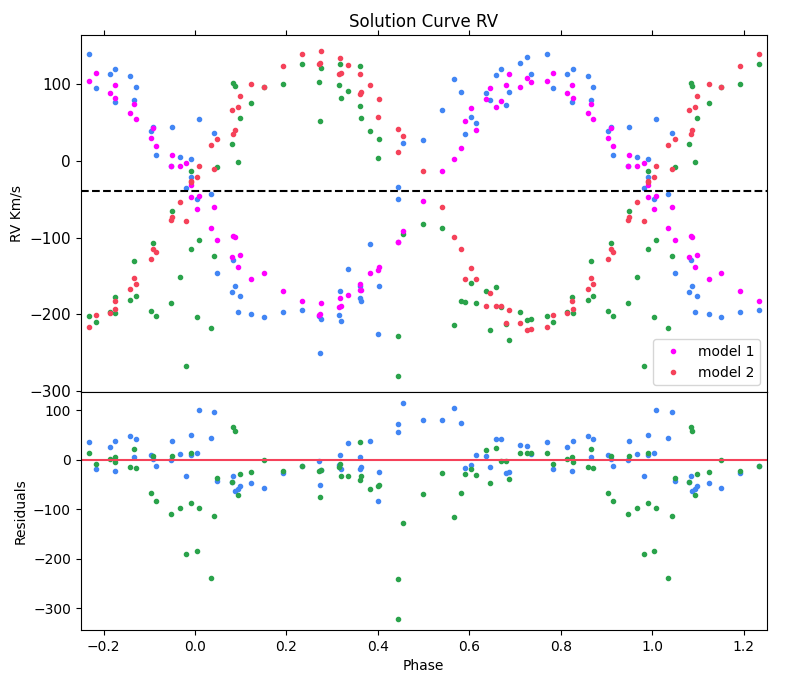}
    \includegraphics[width=\columnwidth]{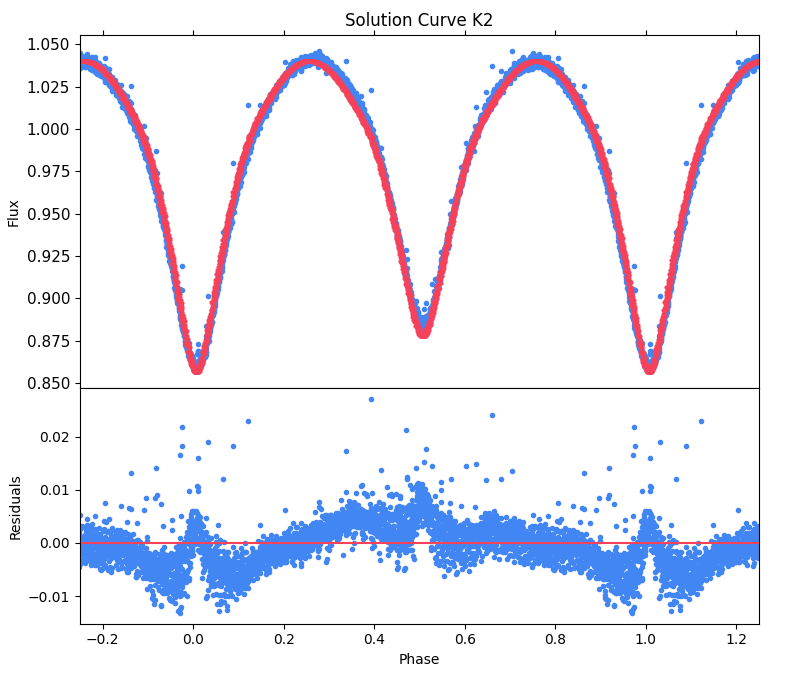}
    \includegraphics[width=\columnwidth]{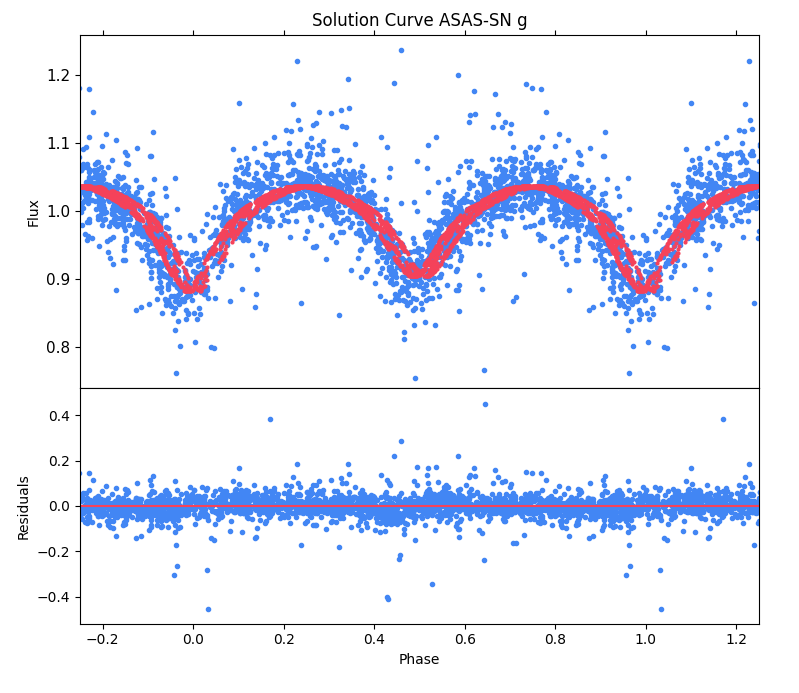}
    \caption{The "toy" model by {\sc W-D} for K2, ASAS-SN $g$ LCs and RV. Top panels show fit of the data, bottom panels show fit residuals. }
    \label{fig:pywd2}
\end{figure}

We use the Wilson–Devinney program {\sc W-D} \citep{wd71,wilson79} with {\sc PYWD2015} \citep{pywd} user interface to make a ``toy" model of LCs from TESS, KEPLER and ASAS-SN together with $\rv_{1,2}$ from our spectral analysis. These datasets were fitted by the differential correction program in contact binary regime ({\sc Mode=3}) with dimensionless potential $\Omega$, inclination $i_{12}$, mass ratio $q_{12}$, semimajor axis $a=a_1+a_2$, conjunction time $t_{0,12}$ as free parameters. For the third  star orbit we assumed the simple case of a circular orbit with a fixed period $P_{\rm 3}=944$ days and  $a_{12} + a_{\rm 3}, t_{0, {\rm 3}}$  as free parameters. The bandpass luminosity $L1$ of the primary component and third light $L3$ were also fitted, while the parameter $L2$ was fixed to 1 by the {\sc W-D}. All other parameters were fixed, including the synchronicity parameters $F_{1,2}=1$ and temperatures. We show the solution derived after 11 iterations in Figure~\ref{fig:pywd2} and Table~\ref{tab:wd}. It is clear that our ``toy" model is unable to properly fit all light variations, especially in K2 and TESS light curves. These LCs are affected by the O'Connell effect \citep{oconnel}, which is caused by spots on stellar surfaces. Residuals have different patterns for different LCs, so spot activity changes with time. ASAS-SN datasets have significantly lower precision and cover a longer time base; therefore, they are less affected by the O'Connell effect. Detailed analysis, including modelling of spots for K2 and TESS LCs will be done in the upcoming paper (Matekov et al in prep). Nevertheless, results of this analysis confirm the presence of the third  star through significant contribution of $L_3$, which gradually decreases from $L3=78\%$ for ASAS-SN $g$ down to $L3=68\%$ for TESS bands. This indicates that the third  star is hotter than components of the contact system, which agrees with the findings from the spectra. Figure~\ref{fig:pywd2} shows that the third  star's attraction causes significant motion to the barycentre of the contact system: RV solution oscillates around the systemic velocity and LC minima show periodic changes due to LTTE, which is clearly visible for ASAS-SN datasets. Therefore, we can perform O-C analysis for the times of minima.

\begin{table}
    \centering
        \caption{The  "toy" model by {\sc W-D}. Error estimates are standard errors from the differential correction program.}
    \begin{tabular}{l|c}
\hline
Parameter & {\sc W-D} value\\
\hline
fixed:\\
$P_{12}$, d  & 0.364277\\
$e_{12}$ & {0.0}\\
$F_{1,2}$ & {1.0}\\
$\alpha_{1,2}$ & 0.32 \\
$A_{1,2}$ &  0.5\\
${\teff}_{1,2}$, K & 5700, 5500\\
$\gamma_{12},\,\kms$ & -38.98\\

$P_{\rm 3}$, d  & 944.0\\
$e_{\rm 3}$ & {0.0}\\
$i_{\rm 3}^{\circ}$  & 90 \\%

\hline
fitted:\\
$a_1+a_2,\,R_\odot$ &  2.468$\pm$0.055 \\
$i_{12}^{\circ}$  & 85.56$\pm$0.12 \\%
$\Omega_{1,2}$ &  $3.458\pm0.002$\\
$q_{12}$ & 0.853$\pm$0.001\\
$t_{0,12}$, BMJD d&  57391.75996$\pm$ 0.00002 \\

$a_{12}+a_{\rm 3},\,R_\odot$ &  563.01$\pm$9.98 \\
$t_{0, {\rm 3}}$, BMJD d&  58033.13$\pm$ 0.52  \\

$L1_g$  &  1.80$\pm$0.03\\
$L1_V$  &  1.96$\pm$0.04\\
$L1_{K2}$  &  2.193$\pm$0.006\\
$L1_{T43}$  &  2.367$\pm$0.007\\
$L1_{T44}$  &  2.317$\pm$0.006\\
$L1_{T70}$  &  2.430$\pm$0.007\\
$L1_{T71}$  &  2.436$\pm$0.007\\
$L3_g$, per cent &  78$\pm$1\\
$L3_V$, per cent &  75$\pm$1\\
$L3_{K2}$, per cent &  71.8$\pm$0.1\\
$L3_{T43}$, per cent &  69.0$\pm$0.1\\
$L3_{T44}$, per cent &  69.6$\pm$0.1\\
$L3_{T70}$, per cent &  68.3$\pm$0.1\\
$L3_{T71}$, per cent &  68.5$\pm$0.1\\

\hline
derived\\%
$M_{1,2},\,M_\odot$ & 0.736, 0.615\\
$R_{1,2},\,R_\odot$ & 0.962, 0.887\\
$\logg_{1,2}$, cgs  & 4.34, 4.33\\%
$\log{L_{1,2}},\,L_\odot$&  -0.058, -0.190 \\
$r_1$ (pole) &  0.3760$\pm$0.0002 \\
$r_1$ (side) &  0.3970$\pm$0.0002 \\
$r_1$ (back) &  0.4305$\pm$0.0003 \\
$r_2$ (pole) &  0.3535$\pm$0.0004 \\
$r_2$ (side) &  0.3725$\pm$0.0005 \\
$r_2$ (back) &  0.4100$\pm$0.0009 \\
\hline
    \end{tabular}
    \label{tab:wd}
\end{table}

\subsection{O-C curve analysis}
\begin{figure}
    \centering
    \includegraphics[width=0.95\columnwidth]{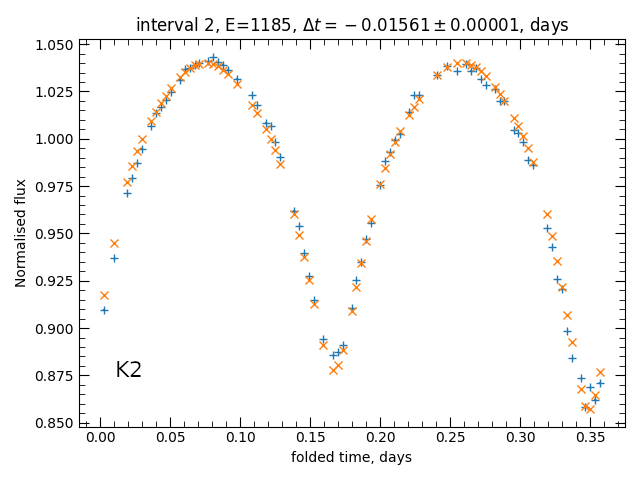}
    \includegraphics[width=0.95\columnwidth]{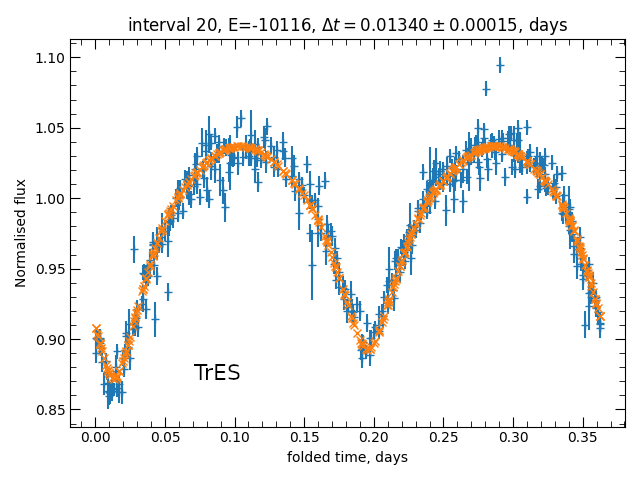}

    \caption{Plots illustrating our template matching algorithm for finding times of minima. Titles indicate duration of time interval, epoch, timeshift and its error. We show it for intervals in  TrES, and K2 datasets. }
    \label{fig:timeshift}
\end{figure}

Unfortunately, our LC datasets mostly have sparse time coverage; therefore, one has to use data from long time intervals to get a good phase coverage of the light curve. Only for TESS data do we have enough time resolution to get times of the minima from every orbital cycle. Thus, we use template matching to get times of the minima, because it allows us to use LCs with sparse phase coverage and allows us to use a distinct shape of LC\citep{cancelKvW}. The linear ephemeris from Equation~\ref{eq:efimer} was used to phase fold all datapoints from the selected time interval. The interval is selected only if it covers at least 90 per cent of the phases and contains more than 19 datapoints. 
We employ our ``toy" model to generate a set of templates for Kepler, TESS, $g$ and $V$ bands. Then we use {\sc scipy.optimise.curve\_fit} function to find the optimal time shift of the relevant template, which is then linearly interpolated to match all datapoints from the interval. We select the integer number closest to the cycle number of the median time for a given interval as an epoch $E$. We show examples of template matching in Figure~\ref{fig:timeshift}. The time shifts and their errors are the same as the primary minima and can be used for the O-C curve analysis. 

\begin{figure*}
    \includegraphics[width=\textwidth]{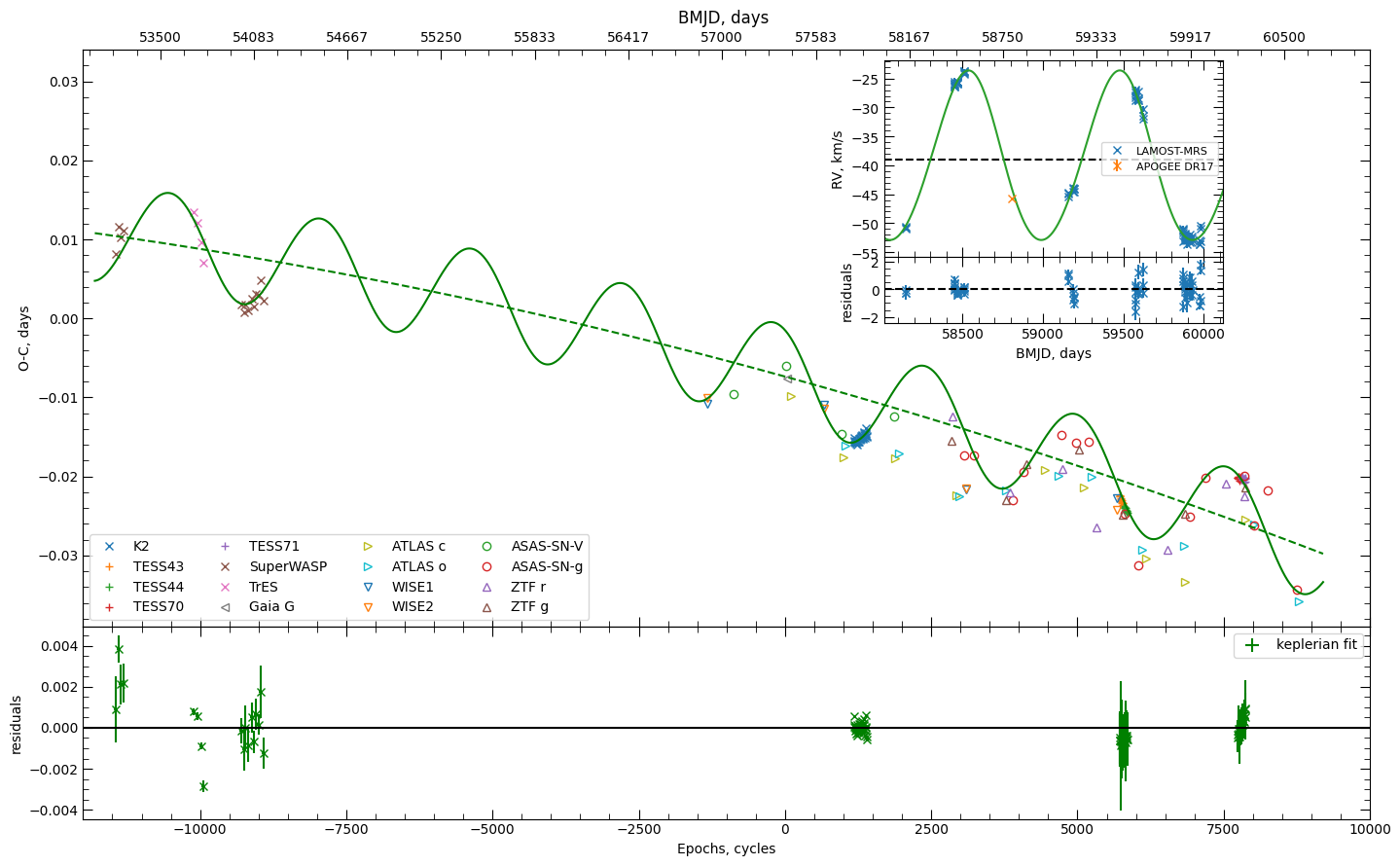}
    \caption{Results for joint fit. O-C curve fit using parabola (dashed line) plus Keplerian orbit (solid line).  Horizontal axes are shown with units in cycles (bottom) and BMJD (top) for convenience. Only K2, TESS, TrES and SuperWASP data are used in fitting, while others are shown for reference. $\rv_3$ curve fit is shown on the inline plot. Dashed horizontal line shows the systemic velocity. $\rv$ measurement from APOGEE DR17 \protect\citep{apogee17} is also shown. Residuals for both fits are shown in the bottom panels.}
    \label{fig:omc}
\end{figure*}

The O-C curve is modelled using parabolic plus periodic variation, in a form of Keplerian orbit, which should take into account the LTTE effect. 

\begin{align}
    O-C=\Delta t_0 + \Delta P~E + \beta~E^2+W(p,E),
\end{align}
where $\Delta t_0,~\Delta P$ are corrections for linear ephemeris, $\beta$ is a long-term change of the period and $W(p,E)$ is a periodic variation, where  $p=A,~P_3,~t_{p3},~e,~\omega$ are parameters of the Keplerian orbit. We used a built-in model from PyAstronomy\citep{pyasl}, which allows the computation of position and velocity along the line of sight for a two-body problem. Therefore, we fit both the O-C and $\rv_3$ datasets simultaneously by adding systemic velocity $\gamma$ and mass ratio $q_3=\frac{M_1+M_2}{M_3}$ to the fitted parameters. Also, we set $\omega^*_3=\omega_3+180^\circ$ and $A=a_{12}\sin{i_3}/c$ with -$c$ speed of light. 
We use only O-C from K2, TESS, TrES and SuperWASP, because they were computed using data from short time intervals ($\Delta T=2$ days for K2, TESS and $\Delta T=20$ days for TrES, SuperWASP), which are significantly smaller than the period ($P_3\sim944$ days) derived from $\rv$ curve.  We sample solutions with {\sc EMCEE} \citep{emcee} using 50 walkers and 25000 iterations.  We present the resulting O-C curve fit in Figure~\ref{fig:omc}, where we also show O-C values from ZTF, WISE, ASAS-SN, Gaia and ATLAS just for reference.

Table~\ref{tab:o-c} lists parameters for joint fit models. We compute final values and their uncertainties as a median and standard deviation of {\sc EMCEE} sampling, shown in Figure~\ref{fig:corner}. This solution is consistent with {\sc GLS} orbit, which was computed earlier, but precision is significantly improved. The contact system is heavier than the third  star with mass ratio $q_3=1.16\pm0.01$. $\beta\neq0$, thus the orbital period of the contact binary decreases at a rate of $dP/dt=-4.29\cdot10^{-8}~{\rm d~yr^{-1}}$.  
The updated linear ephemeris is:
\begin{align}
t_{min}(\rm BJD)= 2457391.90648+0.364275E,
\label{eq:efimer1}
\end{align}

\begin{table}
    \centering
    \caption{Joint $\rv_3$ and O-C curves fit. Estimates and errors are from median and standard deviations of {\sc EMCEE} sampling.}
    \begin{tabular}{lc}
\hline
Parameter & value \\
\hline
$\Delta t_0$, day &  -0.0074$\pm$0.0001 \\
$\Delta P$, $10^{-6}$ day cycle$^{-1}$&  -2.04$\pm$0.01\\
$\beta$, $10^{-11}$ day cycle$^{-2}$  &  -4.28$\pm$0.23\\
$A$, day &    0.00632$\pm$0.00010 \\
$P_3$, day &  941.40$\pm$0.03  \\
$t_{p3}$, (BMJD) day  & 58606.85$\pm$10.10 \\
$e_3$ &  0.059$\pm$0.007\\
${\omega^*_3} ^\circ$  & 210.3$\pm$3.1\\
$q_3=\frac{M_1+M_2}{M_3}$ & $1.16\pm0.01$\\
$\gamma_3,\,\kms$& $-39.00\pm0.06$\\

\hline
    \end{tabular}
    \label{tab:o-c}
\end{table}

\section{SED fitting}
\label{sec:sed}
The spectral energy distribution (SED) offers an independent approach to estimating system parameters like $\teff$ and distance $d$. 
We utilize the {\sc SEDFit} package\footnote{\url{https://github.com/mkounkel/SEDFit} \citep{sedfit}.} for SED fitting, using BT-Settl \citep{btsettl} models to find the optimal fit. We use this software because it allows models with three components, plus it can use Gaia DR3 \citep{gaia3} BP/RP spectrum. Unfortunately, current version doesn't provide uncertainties for fitted parameters. 
We impose constraints on the solution by fixing parameters of the contact system: ${\teff}_{1,2},~\logg_{1,2}$, $R_{1,2}$, $L3_{TESS}=68\%$ from the LC solution (our ``toy" model), while fitting for the distance $d$, extinction $A_V$, $\feh$ and third  star parameters (${\teff}_{3},~\logg_3,~R_3$). 
\par
\begin{figure}
    \includegraphics[width=\columnwidth]{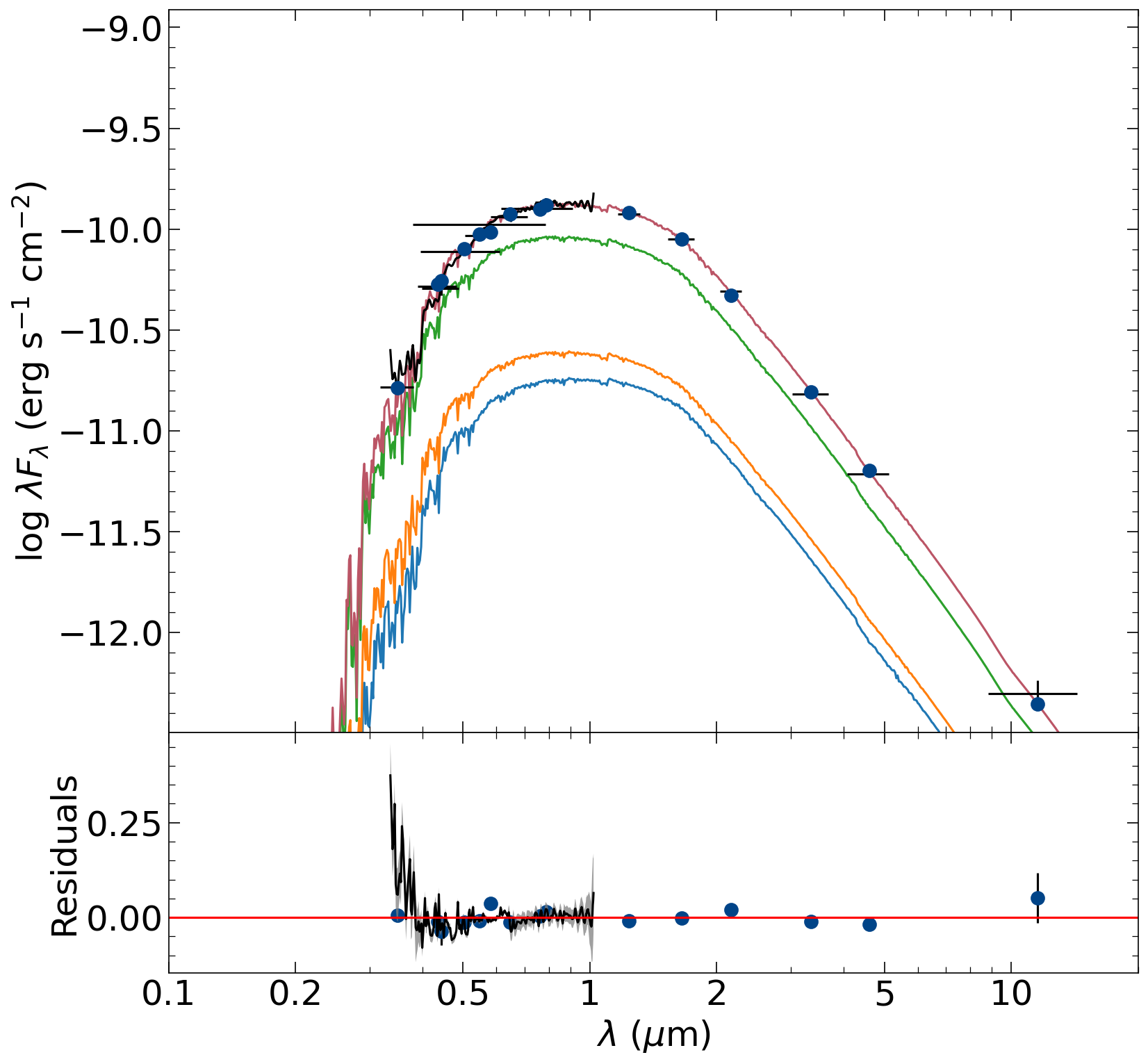}
    \caption{SED fitting with {\sc SEDFit}. The top panel shows the observations, including Gaia DR3 BP/RP spectrum (black line), fluxes in Cousins, Gaia, 2MASS and WISE filters. Best fit SED of the whole system (red line), third  star (green line), primary (orange line) and secondary (blue line) components of the contact system are also shown. The fit residuals are shown on the bottom panel. }
    \label{fig:sed}
\end{figure}

In Figure~\ref{fig:sed}, we present the resulting fit. Photometric measurements did not account for observations taken during eclipses, which may explain the small discrepancy. The estimated distance of $d=619$ pc is larger than the single-star model value of $d=487^{503}_{468}$ pc from Gaia DR3\footnote{GSP-Phot Aeneas best library using BP/RP spectra (distance\_gspphot)}. Other parameters are $A_V=1.18$ mag, $\feh=0.05$ dex, ${\teff}_{3}=5800$ K, $\logg_3=4.31$ dex, $R_3=1.78~R_\odot$. They are quite different from our average spectroscopic values, but similar to Gaia DR3 single-star estimates $\teff=5620_{5595}^{5647}$ K, $\logg=4.12_{4.10}^{4.14}$ dex, [M/H]$=-0.30_{-0.33}^{-0.28}$ dex, $R=1.71_{1.65}^{1.77}~R_\odot$. 

\section{Discussion}
\label{discus}
Results from joint $\rv_3$ \& O-C fitting allow us to get an estimate of the total mass ($M_{\rm tot}=M_1+M_2+M_3$) using Kepler's third law:
\begin{align}
    M_{\rm tot}\sin^3{i_3}=\frac{\left(c~A~\sqrt{1-e_3^2}~(1+q_3)\right)^3}{{GM_{\odot}}} \left[\frac{2 \pi}{P_3}\right]^2=1.95\pm0.03~M_\odot ,
    \label{eq:kepler3}
\end{align}
where $GM_\odot=1.32712440041\cdot 10^{20}\, {\rm m^3\,s^{-2}}$ - is the Solar mass parameter\footnote{\url{https://iau-a3.gitlab.io/NSFA/NSFA_cbe.html\#GMS2012}}. Thus, the contact system and the third star have masses $M_{\rm 12,~3}\sin^3{i_3}=1.05\pm0.02,~0.90\pm0.02~M_\odot$. Combining this with the mass of contact system $M_1+M_2=1.351~M_\odot$ from the ``toy" model, we can get the approximation for $\sin{i_3}=0.9198$, so the inclination angle is $i_3\sim67^\circ$.
\par
Projected semimajor axes for the contact system and the third star are: $a_{12,~3} \sin{i_3}=234\pm3,~272\pm4~R_\odot$. The angular separation between the contact system and the third  star is around $\sim5$ mas, if we use the Gaia DR3 distance $d=486$ pc, which is too small to be resolved. However, photocenter should move relative to the center of mass with amplitude $\sim1$ mas, slightly oscillating due to the variability of the contact system.  Gaia DR3 Non-single star orbital catalog \citep{gaia3nss} has many astrometric orbits with even smaller values of the semimajor axis, although now the astrometric orbit is absent for J04+25. Even if the orbit was computed, it was rejected as dubious by the filter on parallax significance, see Eq. 16 in \citet{gaia3astro}. Possibly, variability of the contact system led to unreliable measurements of the photocenter position (renormalised unit weight error RUWE=$6.029$!), so Gaia was unable to determine a good astrometric orbit.  Also, the current data release is based on observations with a time base covering only one full orbital period. We hope that the upcoming Gaia DR4 will provide an orbital solution for J04+25, so one can get an inclination of the wide orbit.
\par
Third star is bright and close to the contact system, therefore theoretically we should be able to see its light reflected by the contact system. However only extremely precise space-based photometry will be sensible to this effect (reflected light $\sim3$ ppm) which will be seen as a very small increase of $L3$ for illuminated phases.

\subsection{Third  star parameters from APOGEE}
\begin{figure*}
    \includegraphics[width=\textwidth]{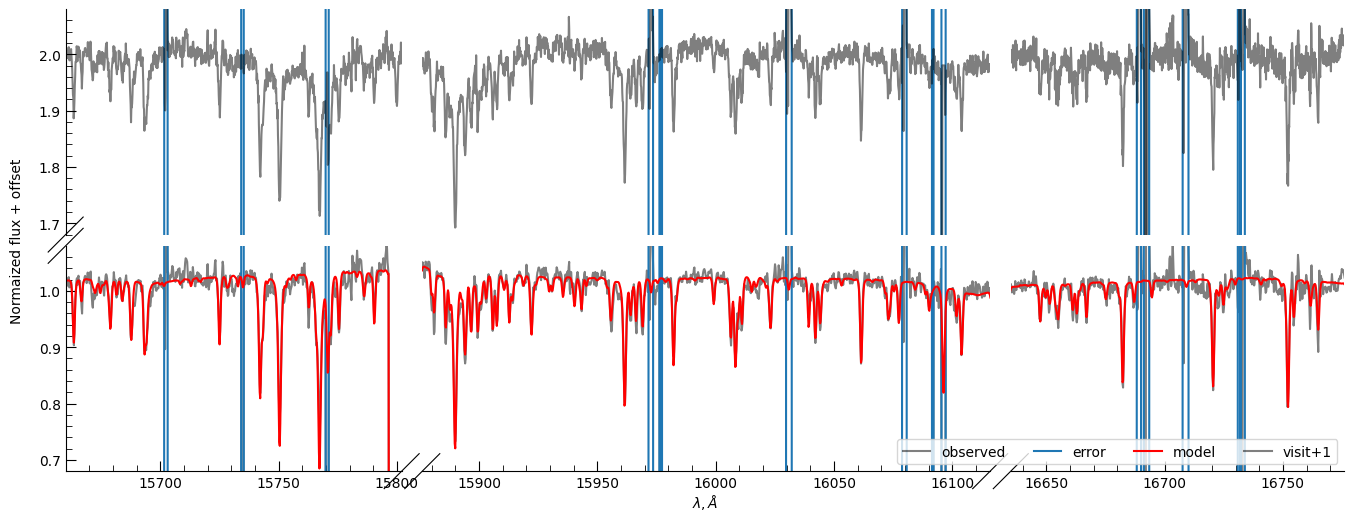}
    \caption{Parts of APOGEE DR17 ASPCAP spectrum (gray line) with best-fit model (red line) from SDSS Science Archive Server are shown on the lower panels. Visit spectrum normalized by the second order polynomial is shown on the top panels}
    \label{fig:apogee}
\end{figure*}
APOGEE DR17 \citep{apogee17} has single-star estimates for this star: $\rv=-45.75 \pm0.05~\kms$, $\teff=6343 \pm 46$~K, $\logg=4.15 \pm 0.03$ dex, [M/H]=$0.05 \pm 0.01$ dex, which corresponds to the third  star. This result is based on high-resolution ($R=\lambda/\Delta\lambda=22500$) infrared spectrum, although it should be biased as the single-star model fit did not take into account the dilution of the light by the contact system, see Figure~\ref{fig:apogee}.
The contact system has nearly 'flat' spectrum due to fast rotation; therefore, the spectral contribution of the contact system can be easily removed during normalization of the original spectrum using a high-order polynomial function. We can see in Figure~\ref{fig:apogee} that the contribution of the contact system was almost completely vanished by the normalisation. Thus we recommend the usage of the low-order polynomial for the normalisation of the spectrum, if we want to keep information about the contact system.

\par

\subsection{GLS fit of $\vsini$}

\begin{figure}
    \includegraphics[width=\columnwidth]{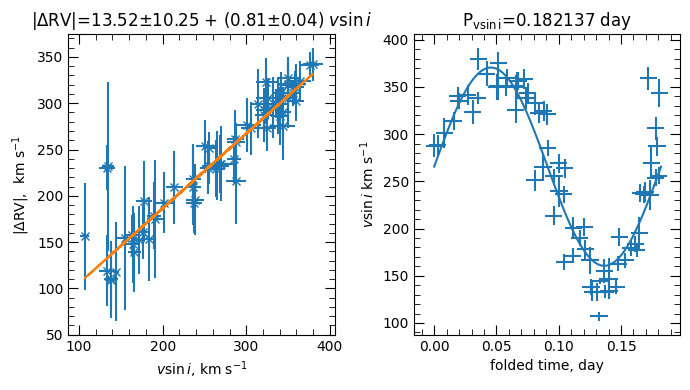}
    \caption{Correlation of $|\Delta\rv|$ with $\vsini$ (left panel). Best sine-fit of $\vsini$ values of whole contact system by {\sc GLS} (right panel).  }
    \label{fig:vsini}
\end{figure}

$\vsini$ values derived by the binary model in the first iteration correlate well with the absolute difference of radial velocities in the contact subsystem (see Figure~\ref{fig:rv12}). Linear fit is shown in the left panel of Figure~\ref{fig:vsini}: 

\begin{align}
    |\Delta\rv|=13.52\pm10.25 + (0.81\pm0.04)\vsini,
\end{align}

It is similar to the empirical relation from \citet{cat22}, but here we probe larger values of $\vsini$: up to $\leq400~\kms$, while \citet{cat22} had $\vsini<300~\kms$. We can try to use this correlation to get an approximation for the sum of radial velocity amplitudes ($K_1+K_2\sim|\Delta\rv|_{\rm max}\sim (\vsini)_{\rm max}$) and the period \citep{j115}. These two values can provide us minimal estimation of the subsystem mass $M_{\rm contact}\sin^3{i}$, using only spectroscopic data. 
\par
We run {\sc GLS }, assuming circular orbit and find the best fit period $P_{\vsini}=0.182137\pm0.000001$ day, which is exactly half of the period from LC, see the right panel in Figure~\ref{fig:vsini}. Amplitude and offset are $K_{\vsini}=105\pm25~\kms$ and $C_0=262\pm5~\kms$ respectively. Using the linear relation we get $(K_1+K_2)=309\pm27~\kms$, which together with doubled period provides us $M_{\rm 12}\sin^3{i_{12}}=1.12\pm0.30~M_\odot$. This value is consistent with $M_{\rm 12}\sin^3{i_3}=1.05\pm0.02~M_\odot$ from the joint fit of the third  star orbit.

\section{Conclusions}
\label{concl}
J04+25 is an interesting case of hierarchical triple, where the third star is significantly brighter than the inner contact subsystem. Partial eclipses in the inner subsystem allow us to estimate third light and determine ``light orbit" through LTTE, while the third  star dominates in the spectra and can provide SB1 orbit. Joint fit of these data gives us complete information on the projection of the wide orbit on the line of sight. Extraction of RVs in the inner subsystem is a complex task for the traditional cross-correlation function (CCF) method because the third  star dominates in the spectrum. Spectral disentangling \citep{fd3} can not be applied here, since light contributions of components change in every spectrum. Therefore, we determine RVs with iterative "Matryoshka"-like approach, using binary spectral model twice. This provides RVs consistent with our expectations for this contact binary.

\par
Here we summarise our results:
\begin{enumerate}
    \item we successfully extracted radial velocities from medium resolution spectra for all three components, using binary spectral model in two steps. Third  star RV have high precision and allow direct fitting of the orbit. Low precision of RV measurements in contact system is compensated by large number statistics. Thus we conclude that medium resolution spectra with low $\snr$ still can be used to study contact binaries,
    \item we built approximate ``toy" model of the inner contact binary with third  star using high quality photometry and RV. Although this model can't properly fit all light variations in space-based LCs, it can serve as a good template for the searching of ``light orbit" through LTTE. More detailed modelling, including spots, will be done in upcoming paper (Matekov et al in prep.),   
    \item we employed template matching technique to find periodic variation in several photometric datasets. This variation is consistent with third  star orbit which was found earlier. Joint fit of RV and O-C curves allowed us to refine third  star orbit.
    \item we made estimations of the masses in wide system and discussed possible determination of astrometric orbit in the future data release of Gaia.
\end{enumerate}

For this object we try to explore all potential of LAMOST-MRS data for the studying of the contact system. We find that even using spectra alone one can get estimations of the orbital period and the sum of radial velocity amplitudes, which taken together with Kepler's third law can provide minimal mass of the contact system. We plan to apply this method for the set of known contact binaries, observed by LAMOST-MRS.


\section*{Acknowledgements}
 We are grateful to our referee, Andrei Tokovinin, for a constructive report. His suggestions have significantly improved this article.
\par

Guoshoujing Telescope (the Large Sky Area Multi-Object Fiber Spectroscopic Telescope LAMOST) is a National Major Scientific Project built by the Chinese Academy of Sciences. Funding for the project has been provided by the National Development and Reform Commission. LAMOST is operated and managed by the National Astronomical Observatories, Chinese Academy of Sciences. 
This work is supported by the Natural Science Foundation of China (grant Nos. 12288102, 12090040/3, 12473034, 12103086, 12273105, 11703081, 11422324, 12073070), the National Key R$\&$D Program of China (grant Nos. 2021YFA1600403, 2021YFA1600400), CAS project XDB1160000/XDB110201, Yunnan Fundamental Research Projects (Nos. 202401BC070007 and 202201BC070003), the Yunnan Revitalization Talent Support Program–Science $\&$ Technology Champion Project (No. 202305AB350003), the Key Research Program of Frontier Sciences of CAS (No. ZDBS-LY-7005), Yunnan Fundamental Research Projects (grant Nos. 202301AT070314, 202101AU070276, 202101AV070001), and the International Centre of Supernovae, Yunnan Key Laboratory (No. 202302AN360001). We also acknowledge the science research grant from the China Manned Space Project with Nos. CMS-CSST-2021-A10 and CMS-CSST-2021-A08. The authors gratefully acknowledge the “PHOENIX Supercomputing Platform” jointly operated by the Binary Population Synthesis Group and the Stellar Astrophysics Group at Yunnan Observatories, Chinese Academy of Sciences. 
This research has made use of NASA’s Astrophysics Data System, the SIMBAD data base, and the VizieR catalogue access tool, operated at CDS, Strasbourg, France. It also made use of TOPCAT, an interactive graphical viewer and editor for tabular data \citep[][]{topcat}. 
Funding for the TESS mission is provided by NASA’s Science Mission directorate. This paper includes data collected by the TESS mission, which is publicly available from the Mikulski Archive for Space Telescopes (MAST). 
This work has made use of data from the European Space Agency (ESA) mission {\it Gaia} (\url{https://www.cosmos.esa.int/gaia}), processed by the {\it Gaia} Data Processing and Analysis Consortium (DPAC, \url{https://www.cosmos.esa.int/web/gaia/dpac/consortium}). Funding for the DPAC has been provided by national institutions, in particular the institutions participating in the {\it Gaia} Multilateral Agreement.
This research has made use of the NASA/IPAC Infrared Science Archive, which is funded by the National Aeronautics and Space Administration and operated by the California Institute of Technology.

\section*{Data Availability}
The data underlying this article will be shared on reasonable request to the corresponding author.




\bibliographystyle{mnras}




\appendix

\section{Spectral models}
\label{sec:payne}
The synthetic spectra are generated using NLTE~MPIA online-interface \url{https://nlte.mpia.de} \citep[see Chapter~4 in][]{disser} on wavelength intervals 4870:5430 \AA~for the blue arm and 6200:6900 \AA ~for the red arm with spectral resolution $R=7500$. We use NLTE (non-local thermodynamic equilibrium) spectral synthesis for H, Mg~I, Si~I, Ca~I, Ti~I, Fe~I and Fe~II lines \citep[see Chapter~4 in][ for references]{disser}.  
\par
The grid of models (6200 in total) is computed for points randomly selected in a range of $\teff$ between 4600 and 8800 K, $\logg$ between 1.0 and 4.8  (cgs units), $\vsini$ from 1 to 300 $\kms$ and [Fe/H]\footnote{We used $\feh$ as a proxy of overall metallicity, abundances for all elements are scaled with Fe.} between $-$0.9 and $+$0.9 dex. The model is computed only if linear interpolation of the MAFAGS-OS\citep[][]{Grupp2004a,Grupp2004b} stellar atmosphere is possible for a given point in parameter space.  Microturbulence is fixed to $\Vmic=2~\kms$ for all models. 

\section{RV \& O-C measurements }
 We provide the RV measurements in Tables~\ref{tab:rv3},~\ref{tab:rv12}. O-C values are listed in Table~\ref{tab:minima}.
\begin{table}
    \centering
    \caption{\label{tab:rv3} Radial velocity measurements for third star. Full table is available online}
    \begin{tabular}{lcc}
\hline
time BMJD &  value & error  \\
 day    & $\kms$ & $\kms$\\

\hline
58143.4707 &-50.8668 &0.4356\\
58143.4874 &-50.6238 &0.3556\\
58143.5033 &-50.8299 &0.3556\\
.. & .. & ..\\
 \hline
\end{tabular}
\end{table}

\begin{table}
    \centering
    \caption{\label{tab:rv12} Radial velocity measurements for components in the contact subsystem. Full table is available online.}
    \begin{tabular}{lcc}
\hline
time BMJD & RV$_{1}$ & RV$_{2}$ \\
 day  & $\kms$ & $\kms$  \\
 \hline
58143.4874 &49.9$\pm$ 24.0 &-184.7 $\pm$16.5\\
58143.5033 &111.9$\pm$ 14.3 &-164.8 $\pm$15.0\\
58450.6503 &119.7$\pm$ 8.3 &-197.8 $\pm$9.5\\
.. & .. & ..\\
 \hline
\end{tabular}
\end{table}

\begin{table}
    \centering
    \caption{\label{tab:minima} O-C values. Full table is available online.}
    \begin{tabular}{lccc}
\hline
LC & cycle & interval & timeshift (O-C) \\
& & day  & day   \\
 \hline
K2 &1180 &2 &-0.01510 $\pm$0.00001\\
K2 &1185 &2 &-0.01561 $\pm$0.00001\\
K2 &1191 &2 &-0.01549 $\pm$0.00001\\
K2 &1196 &2 &-0.01577 $\pm$0.00001\\
.. & .. & .. & ..\\
 \hline
\end{tabular}
\end{table}

\section{{\sc emcee} sample of the joint O-C\&RV solution}

\begin{figure*}
    \includegraphics[width=\textwidth]{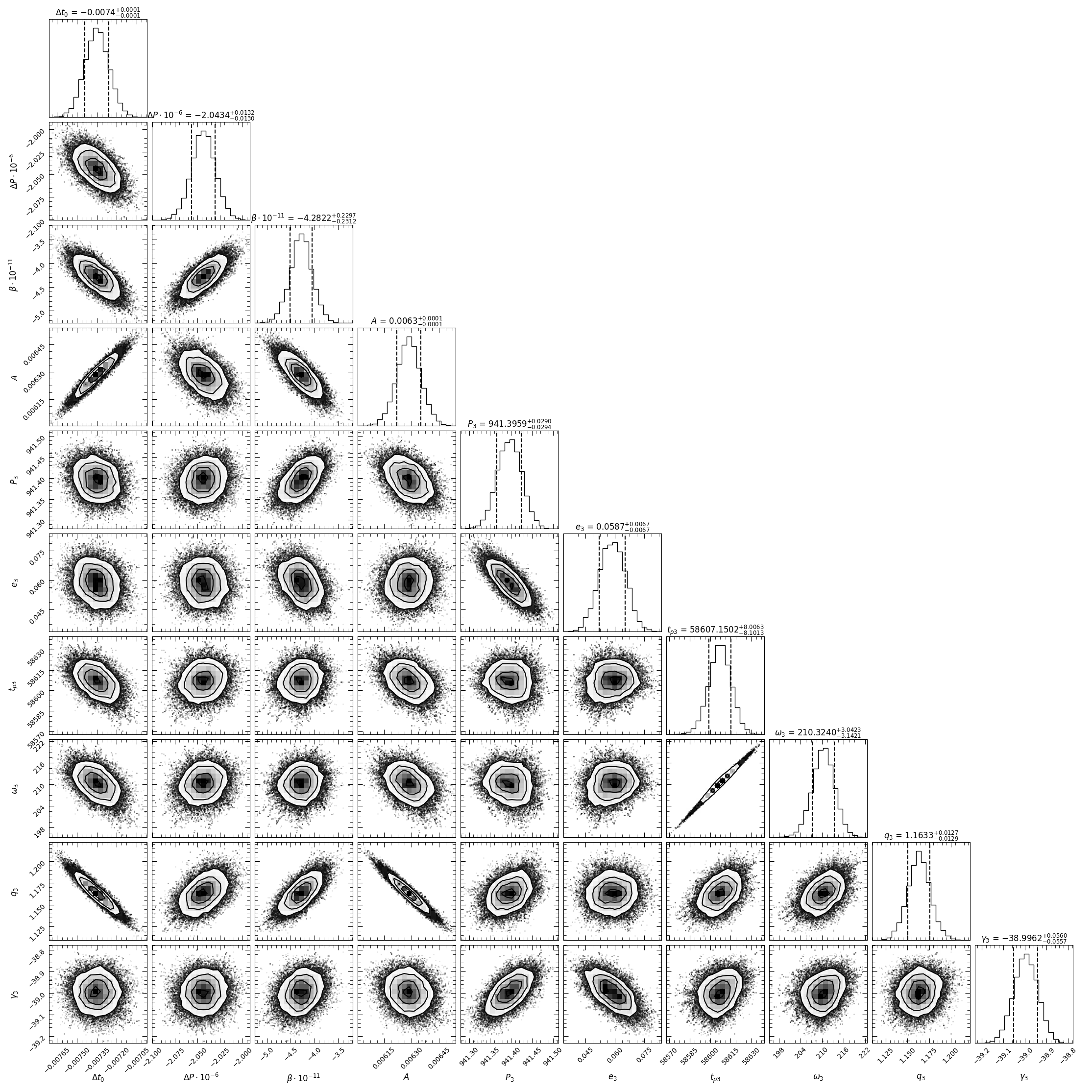}
    \caption{Corner plot for the sampling of the joint solution. Titles show 16, 50 and 84 percentiles.}
    \label{fig:corner}
\end{figure*}

In Figure~\ref{fig:corner} we show corner plot \citep{corner} with {\sc emcee} sampling results for the joint O-C\&RV solution. We removed results in chains for four walkers, which got stuck in wrong periods.

\bsp	
\label{lastpage}
\end{document}